\documentstyle[12pt]{article}

\begin{document}

\begin{center}

{\Large{\bf Self--Similar Perturbation Theory} \\ [5mm]

V. I. Yukalov$^{1,2}$ and E. P. Yukalova$^{2,3}$} \\ [2mm]

{\it $^1$Bogolubov Laboratory of Theoretical Physics \\
Joint Institute for Nuclear Research, Dubna 141980, Russia \\

\vskip 2mm

$^2$Instituto de Fisica de S\~ao Carlos, Universidade de S\~ao Paulo \\
Caixa Postal 369, S\~ao Carlos, S\~ao Paulo 13560--970, Brazil \\

\vskip 2mm

$^3$Department of Computational Physics \\
Laboratory of Computing Techniques and Automation \\
Joint Institute for Nuclear Research, Dubna 141980, Russia}

\end{center}

\vskip 1.5cm

\begin{abstract}

A method is suggested for treating those complicated physical problems 
for which exact solutions are not known but a few approximation terms of 
a calculational algorithm can be derived. The method permits one to 
answer the following rather delicate questions: What can be said about 
the convergence of the calculational procedure when only a few its terms
are available and how to decide which of the initial approximations of the 
perturbative algorithm is better, when several such initial approximations 
are possible? Definite answers to these important questions become 
possible by employing the self--similar perturbation theory. The novelty of
this paper is in developing the stability analysis based on the method of
multipliers and in illustrating the efficiency of this analysis by different
quantum--mechanical problems.

\end{abstract}

\vskip 2cm


\newpage

Running head: Self--Similar Perturbation Theory

\vskip 5mm

Number of pages: 45

\vskip 2mm

Number of Tables: 2

\vskip 2cm

{\large{\bf Mailing addresses:}}

\vskip 3mm

{\bf till October 10, 1999}:

\vskip 3mm

Prof. V.I. Yukalov

\vskip 2mm

Instituto de Fisica de Sao Carlos

Universidade de Sao Paulo

Caixa Postal 369, Sao Carlos

Sao Paulo 13560-970, Brazil

\vskip 3mm

Phone:  55 (16) 271-2012

Fax:    55 (16) 271-3616

E-mail: yukalov@if.sc.usp.br

\vskip 5mm

{\bf after October 10, 1999}:

\vskip 3mm

Prof. V.I. Yukalov

\vskip 2mm

Bogolubov Laboratory of Theoretical Physics

Joint Institute for Nuclear Research

Dubna 141980, Russia

\vskip 2mm

E-mail: yukalov@thsun1.jinr.ru

\vskip 2mm

Tel: 7 (096) 216-3947 (office); 7 (096) 213-3824 (home)

\vskip 2mm

Fax: 7 (096) 216-5084

\newpage

\section{Introduction}

Realistic physical problems can practically never be solved exactly, 
quantum field theory being among these the most difficult problems. Then 
one has to resort to some approximations. One such a reliable 
approximation is the Gaussian--effective--potential approach [1--5]. This 
approach contains the leading $1/N$ result as its formal 
$N\rightarrow\infty$ limit and also the one--loop result as its formal 
$\hbar\rightarrow 0$ limit. Thus, it encompasses both of the other 
popular approaches to effective potentials, transforming their 
leading--order expansions and displaying a much richer structure [5,6]. 
At zero temperature, the Gaussian effective potential can be described as 
a variational approximation to the vacuum energy density constructed with 
the Gaussian trial wave functions shifted by a constant classical 
background field [7]. The finite--temperature Gaussian effective 
potential is just the approximate free energy in the self--consistent 
harmonic approximation [7,8]. In this way, the Gaussian effective 
potential in quantum field theory is a direct analog of the variational 
ground--state energy in quantum mechanics, so that the 
quantum--mechanical language can be directly transcribed to the 
quantum--field--theory language [1]. The difference is that variational 
methods in quantum mechanics permit one to derive several nice 
comparison theorems [9--13] for the ground--state energy, which is rather 
difficult, if possible, in quantum field theory.

To obtain corrections to a chosen initial approximation, one has to 
develop some perturbative algorithm about the given approximation. Thus 
post--Gaussian corrections to the Gaussian effective potential may be 
obtained [14,15], allowing the variational parameter to change from one 
order to the next for the expansion to yield convergent results [14--17]. 
In the same spirit, one can calculate the effective potentials or energies 
as well as wave functions [18,19]. Different variants of such an approach 
are called in literature by various names, as modified perturbation theory, 
renormalized perturbation theory, optimized perturbation theory, controlled 
perturbation theory, oscillator--representation method, and so on [20--31].
One often dignifies this kind of methods as nonperturbative, implying that
they result in modified, or resummed, expansions having a more complicated 
structure than simple weak--coupling power series. However, the term 
"nonperturbative" is, to our mind, not only  cumbersome but somewhat 
misleading. This is because from the mathematical point of view any regular
procedure producing a sequence of approximations is {\it perturbation
theory}, irrespectively of what initial approximation is chosen and what 
additional conditions in the course of calculation are imposed. The role 
of such additional conditions is to reorganize, by defining control 
functions, a given series into another one with better convergence 
properties [20--22], which is necessary because of the divergence of 
standard weak--coupling power series [32]. Note that the renormalization 
scale, appearing in perturbative expansions for observable quantities of 
field theories, can also be treated as a control function that may be 
defined by some additional conditions [27,28,33] or invoking information 
from experiment [33,34].

The convergence of modified perturbation theory can be treated in three
ways. The straightforward case is when the model considered is simple 
enough to permit the evaluation of perturbative terms of arbitrary large 
orders. This happens for zero-- and one--dimensional anharmonic 
oscillators [35--37]. But clearly one cannot expect to have the same luck 
for less trivial models.

Another possibility to check convergence is when the exact or very 
accurate numerical solution of the problem is available. This is so for 
many quantum--mechanical problems. Then one may directly compare the 
calculated perturbative terms with the known accurate solutions. If these 
terms approach the latter, one tells that the perturbation procedure is 
convergent. Since in this case one deals with a finite number of 
approximate terms, it is more correct to say that one observes {\it 
numerical convergence} or {\it local convergence}. It is this type of 
convergence that one deals with in any calculational procedure stopping 
at a finite--order approximation. Certainly, the luxury of possessing for 
comparison exact or almost exact numerical results is rather rare for 
realistic physical problems.

The numerical or local convergence may be also observed when one is able 
to find quite a number of perturbative terms, say several tens or hundreds 
of them, so that the calculational procedure saturates at some value that 
changes less and less with the increasing order of approximation. Again, 
it is not too realistic to hope to reach such a saturation with tens or 
hundreds of perturbative terms in complicated problems, such as field
theory, usually allowing just a few terms.

What then can be said about the reliability of the results obtained by 
means of a perturbation algorithm, even if this is an optimized or 
modified perturbation theory, when one has in hands only a few terms and no 
exact answers are known? It may happen that the subsequent perturbative 
results quantitatively are not close to each other, then which of them is 
preferable? For example, optimized perturbation theory has been 
systematically tested using different ansatze for a range of different 
physical quantities in quantum electrodynamics and quantum chromodynamics 
[33]. Although in some specific applications this theory has met with 
success, in the majority of tests the optimization procedure was not 
successful. Typically, though the sign of the coefficients are correctly 
predicted, the optimization ansatze give values that are an order of 
magnitude too large or too small. Thus one comes to the conclusion that 
it is difficult to be optimistic about the general usefulness of 
optimized perturbation theory [33]. Even the more dramatic situation is 
when the neighbour--order results differ qualitatively, thus yielding 
different physics. For instance, for $(2+1)$--dimensional scalar theory, 
the analysis of the Gaussian effective potential shows a first--order 
phase transition, in contrast with the post--Gaussian approximation 
indicating a phase transition of second order [15]. Another example is 
the so--called autonomous version of the scalar field theory, which 
exists in the Gaussian approximation but does not survive, becoming 
unstable, in the higher orders of the post--Gaussian expansion [15].

Thus, for any calculational algorithm, including all variants of modified 
or optimized perturbation theory, there exists the general question: "How 
can one trust to a perturbative procedure when only a few first 
approximations are available and neither exact solutions, nor accurate 
numerical calculations, nor detailed experiments are known for comparison?"
To say this in other words: "Is it possible to formulate a calculational 
algorithm that would be equipped by an internal self--consistent criterion 
allowing to estimate the validity of the obtained approximations?" In the 
present paper we aim at suggesting a positive answer to this question. The
novelty of this paper is twofold: First, we develop the stability analysis,
based on the method of multipliers, so that it really becomes possible to
check the validity of the calculated approximations. Second, we illustrate
the efficiency of the developed analysis by a number of quantum--mechanical
problems. Among the latter, we consider the zero--dimensional $\varphi^4$
theory and anharmonic oscillators with integer powers. Although these models
have been treated earlier by the self--similar approximation theory, however
the stability analysis, as developed in this paper, have not been applied
to them. The main part of the paper is devoted to the problems that have not
yet been treated by the self--similar perturbation theory. These are the
Hamiltonians with power--law potentials, having arbitrary noninteger powers,
and a Hamiltonian with a logarithmic potential. In this way, the results
presented in the paper are new.

\section{Basic Formulas}

Our consideration is based on the self--similar approximation theory 
[38--45]. This approach is more general than modified perturbation theory 
because of the possibility to reformulate calculational procedure as the 
problem of analyzing the evolution equation for a controlled dynamical 
system, thus, permitting us to employ powerful techniques of dynamical 
theory and optimal control theory. We shall not repeat here the mathematical
foundations for the self--similar approximation theory, which have been
expounded in detail in earlier papers [38--45]. We will only mention the
main idea of the approach and delineate its scheme, necessary for the
following illustrations; but we shall concentrate on the {\it stability
problem} that has not yet been considered carefully, the problem which
is pivotal for the motivation of this paper, and which would make it 
possible to answer the questions formulated in the Introduction.

Assume that we are looking for a function $f(x)$ of a variable $x$. For 
simplicity, we imply that the function and variable are real, though the 
whole procedure can be straightforwardly extended to the case of complex 
functions and variables. Let the sought function $f(x)$ cannot be found 
exactly but only its perturbative expressions $p_k(x)$, with 
$k=0,1,2,\ldots$, can be obtained in the asymptotic vicinity of some 
point, say $x=0$. That is,
\begin{equation}
\label{1}
f(x) \simeq p_k(x) \qquad (x\rightarrow 0) \; .
\end{equation}
The standard form of $p_k(x)$ is a series in powers of $x$, because of 
which $x$ is called the {\it expansion parameter}. This can be, e.g., a 
coupling parameter. Such series are practically always divergent. If a 
number of terms $p_k(x)$, about tens or more, were known, one could 
invoke some resummation techniques for ascribing a meaningful value to a 
divergent series [46]. But these techniques are useless when only a few 
perturbative terms are available. With the knowledge of just a few terms, 
one could find the limit of a sequence if a recurrence relation between 
subsequent terms were known. However, such recurrent relations are not 
merely difficult to discover but they have sense only for convergent 
sequences. Really, for an asymptotic series $p_k(x)$, meaningful solely 
for asymptotically small $x\rightarrow 0$ and diverging for any finite 
$x$, a recurrence relation between the terms $p_k(x)$ and $p_{k+1}(x)$ 
cannot be defined since the limit of $p_k(x)$, as $k\rightarrow\infty$, 
does not exist. Nevertheless, if the sequence of approximations 
$\{ p_k(x)\}$ is obtained from one given system of equations, 
corresponding to a physical problem under consideration, by means of 
the same perturbative algorithm, the set of found terms $p_k(x)$ 
does contain information about the sought function $f(x)$. But this 
information, because of divergence of the sequence $\{ p_k(x)\}$, is hidden, 
or we can say that it is enciphered, encoded.

The first thing we have to do in order to decipher, to decode the hidden 
information is to reorganize the sequence $\{ p_k(x)\}$ to another 
sequence that would be convergent at finite $x$. In the process of this 
reorganization, or renormalization, that can be symbolically denoted as
\begin{equation}
\label{2}
{\cal U}\{ p_k(x)\} = \{ F_k(x,u)\} \; ,
\end{equation}
we introduce auxiliary functions $u=u_k(x)$ so that the sequence $\{ 
f_k(x)\}$ of the terms
\begin{equation}
\label{3}
f_k(x) \equiv F_k(x,u_k(x))
\end{equation}
be convergent [20]. Because of the role of the functions $u_k(x)$ in 
controlling convergence, they can be named the {\it control functions}. 
For a convergent sequence $\{ f_k(x)\}$ a limit $f^*(x)$, as 
$k\rightarrow\infty$, exists, which represents the sought function $f(x)$. 
Now the problem of finding a recurrence relation, or mapping, between 
subsequent terms $f_k(x)$ and $f_{k+1}(x)$ can be meaningful.

The main idea in finding a relation between the terms of a sequence 
$\{ f_k(x)\}$ is to try to construct a dynamical system with discrete 
time, being the approximant number $k=0,1,2,\ldots$, so that the motion 
from $f_k(x)$ to $f_{k+n}(x)$ would correspond to the evolution of this 
dynamical system. For this purpose, we need, first, to pass from the 
sequence $\{ f_k(x)\}$ to a set of endomorphisms, which is done as 
follows. Define the {\it expansion function} $x_k(\varphi)$ by the equation
\begin{equation}
\label{4}
F_0(x,u_k(x)) =\varphi\; , \qquad x=x_k(\varphi) \; .
\end{equation}
Introduce an endomorphism $y_k$ by the transformation
\begin{equation}
\label{5}
y_k(\varphi) \equiv f_k(x_k(\varphi)) \; ,
\end{equation}
where, by definition (4), $y_0(\varphi)=\varphi$. As is evident, the 
sequences $\{ y_k(\varphi)\}$ and $\{ f_k(x)\}$ are bijective, and the 
limit $y^*(\varphi)$ of the former sequence is in one--to--one correspondence
with the limit $f^*(x)$ of the latter sequence. The limit $y^*(\varphi)$ 
of the sequence of endomorphisms $\{ y_k(\varphi)\}$ is nothing but a 
fixed point for which
\begin{equation}
\label{6}
y_k(y^*(\varphi)) = y^*(\varphi) \; .
\end{equation}
A family of endomorphisms in the vicinity of a fixed point satisfies the 
equation
\begin{equation}
\label{7}
y_{k+n}(\varphi) = y_k(y_n(\varphi)) \; ,
\end{equation}
which can be easily shown by checking that, for $\varphi\rightarrow y^*$, 
Eq. (7) reduces to the identity $y^*=y^*$. In physical parlance, Eq. (7)
is termed the property of {\it self--similarity}, the usage of which 
justifies the adjective self--similar, as is employed with respect to 
perturbation theory in the title of this paper. In the mathematical 
language, the equations $y_{k+n}=y_k\cdot y_n$ and $y_0=1$ are the 
semigroup properties. A family of endomorphisms $\{ y_k|\; k=0,1,2,\ldots\}$, 
equipped with the semigroup properties, forms a cascade, that is, a 
dynamical system with discrete time. Since, by construction, the cascade 
trajectory $\{ y_k(\varphi)\}$ is bijective to the approximation 
sequence $\{ f_k(x)\}$, we call $\{ y_k|\; k=0,1,2,\ldots\}$ the {\it 
approximation cascade}.

To consider time evolution, it is convenient to deal with continuous 
time. For this, we may embed the cascade into a flow,
\begin{equation}
\label{8}
\{ y_k|\; k=0,1,2,\ldots\} \in \{ y_t| \; t\geq 0\} \; ,
\end{equation}
which implies that the flow trajectory passes through all points of the 
cascade trajectory,
\begin{equation}
\label{9}
y_t(\varphi)=y_k(\varphi) \qquad ( t=k=0,1,2,\ldots) \; ,
\end{equation}
and the same semigroup property, as for the cascade, is true for the flow,
\begin{equation}
\label{10}
y_{t+\tau}(\varphi)  = y_t(y_\tau(\varphi)) \; .
\end{equation}
The flow $\{ y_t|\; t\geq 0\}$ embedding the approximation cascade is 
called the {\it approximation flow}.

From the group relation (10), the Lie equation
\begin{equation}
\label{11}
\frac{\partial}{\partial t} y_t(\varphi) = v(y_t(\varphi))
\end{equation}
immediately follows, in which $v(y)$ is a velocity field. We need to find 
a fixed point $y^*(\varphi)$ of the approximation flow whose motion is 
given by the evolution equation (11). Consider the latter starting from 
the point $y_k(\varphi)$ at $t=k$ and moving to an approximate expression 
for the fixed point $y^*_{k+1}(\varphi,\tau)\equiv y^*_{k+\tau}(\varphi)$
occurring at the time $t=k+\tau$. Then, integrating Eq. (11), we come to 
the {\it evolution integral}
\begin{equation}
\label{12}
\int_{y_k}^{y_{k+1}^*} \frac{dy}{v(y)} = \tau \; ,
\end{equation}
in which $y_{k+1}^* = y_{k+1}^*(\varphi,\tau)$ and $y_k=y_k(\varphi)$.

To make the evolution integral (12) useful, we have to concretize the 
flow velocity. The latter can be done by the {\it Euler discretization}
giving for the velocity $v(y_k(\varphi))\equiv v_k(\varphi)$ the form
\begin{equation}
\label{13}
v_k(\varphi) = F_{k+1}(x,u_k) - F_k(x,u_k) + ( u_{k+1} - u_k)
\frac{\partial}{\partial u_k} F_k(x,u_k) \; ,
\end{equation}
where $x=x_k(\varphi)$ and $u_k=u_k(x_k(\varphi))$. The discretized
velocity (13) may be termed the {\it cascade velocity}. To obtain an 
explicit expression for the latter, we need yet to define the control 
functions $u_k(x)$. This can be realized being based on the following 
reasoning. As is clear from the Lie evolution equation (11), at the fixed 
point $y^*$, when $y_t(y^*)=y^*$, the flow velocity is zero, $v(y^*)=0$. 
Hence, closer we are to a fixed point, smaller is the absolute value of 
velocity. This suggests that the fastest convergence to the fixed point 
is achieved if control functions provide the minimum of the velocity modulus
\begin{equation}
\label{14}
| v_k(\varphi)| \leq | F_{k+1}(x,u_k) - F_k(x,u_k) | +
\left | (u_{k+1} - u_k)
\frac{\partial}{\partial u_k} F_k(x,u_k)\right | \; .
\end{equation}
To minimize the second term in the right--hand side of the latter inequality,
we may set
\begin{equation}
\label{15}
( u_{k+1} - u_k ) \frac{\partial}{\partial u_k} F_k(x,u_k) = 0 \; .
\end{equation}
The principal usage of Eq. (15) is as follows. We are looking for a
solution of the equation
$$
\frac{\partial}{\partial u_k} F_k(x,u_k) = 0 \; , \qquad
u_{k+1}\neq u_k \; ,
$$
which is labelled as the principal of minimal sensitivity [27], and which 
defines the control function $u_k(x)$. If the latter equation has no 
solution, we put $u_{k+1}=u_k$. If this equation possesses several 
physically meaningful solutions, we must take that of them which 
minimizes the cascade velocity (14). In the other way, we could set to zero 
the first term in the right--hand side of Eq. (14), thus coming to the 
principle of minimal difference [20]. But then we should invoke some 
additional condition for defining $u_{k+1}$. In what follows, we use the 
fixed--point condition (15), so that the cascade velocity (13) becomes
\begin{equation}
\label{16}
v_k(\varphi) = F_{k+1}(x,u_k) - F_k(x,u_k) \; ,
\end{equation}
where $x=x_k(\varphi)$ and $u_k=u_k(x_k(\varphi))$.

In this way, taking into account Eqs. (3)--(5), we rewrite the evolution 
integral (12) in the form
\begin{equation}
\label{17}
\int_{f_k}^{f_{k+1}^*} \frac{d\varphi}{v_k(\varphi)} = \tau \; ,
\end{equation}
in which $f_k^*=f_k^*(x,\tau)$ is a $k$--order {\it self--similar 
approximant} for the sought function $f(x)$ and $f_k=f_k(x)$. The 
parameter $\tau$ in Eq. (12) is an effective minimal time necessary for 
reaching a fixed point $y_{k+1}^*$. If no constraints are imposed on the 
properties of the sought function, then the minimal time is the minimal 
number of steps, that is $\tau=1$. When some additional conditions are 
imposed, which the sought function has to satisfy, then $\tau=\tau_k(x)$ 
is to be treated as another type of control functions defined by these 
conditions. One more possibility of fixing $\tau$ is connected with the 
problem of stability, as is described below.

The {\it problem of stability} is extremely important. This is just what 
permits us to decide which of the found self--similar approximants are
trustworthy and which of them are better then others. Let a set of 
mappings $\{ y_k(\varphi)\}$ be given. Stability analysis is based on the 
concept of the {\it local mapping multipliers}
\begin{equation}
\label{18}
\mu_k(\varphi) \equiv \frac{\partial}{\partial\varphi} y_k(\varphi) \; .
\end{equation}
When
\begin{equation}
\label{19}
|\mu_k(\varphi) | < 1
\end{equation}
for a fixed $k$, one says that the mapping at a $k$--step is locally 
stable with respect to the initial point $\varphi$. If $|\mu_k(\varphi)|=1$,
one tells that the mapping is locally neutral. In our case, the sequence 
$\{ y_k(\varphi)\}$ is the trajectory of the approximation cascade. 
Therefore, if inequality (19) holds true, we should tell that the cascade 
at a $k$--step, or the trajectory at a $k$--point, is locally stable with 
respect to the initial point $\varphi$. Similarly, by means of the 
{\it ultralocal multiplier}
\begin{equation}
\label{20}
\overline\mu_k(\varphi) \equiv 
\frac{\delta y_k(\varphi)}{\delta y_{k-1}(\varphi)} =
\frac{\mu_k(\varphi)}{\mu_{k-1}(\varphi)}
\end{equation}
one may characterize the local stability with respect to the variation of the
value $y_{k-1}$ preceeding the point $y_k$, which takes place for
\begin{equation}
\label{21}
|\overline\mu_k(\varphi)| < 1 \; .
\end{equation}
For brevity, we shall say that the cascade at a $k$--step is locally 
stable if inequality (19) is satisfied and it is ultralocally stable when 
condition (21) is valid. The latter can also be called the contraction 
condition. The images of the corresponding multipliers for the domain of 
the variable $x$ are defined as
\begin{equation}
\label{22}
m_k(x) \equiv \mu_k(F_0(x,u_k(x))) \; , \qquad
\overline m_k(x) \equiv\overline\mu_k(F_0(x,u_k(x))) \; ,
\end{equation}
in accordance with Eq. (4). Note that the local multipliers (18) define the
local Lyapunov exponents
\begin{equation}
\label{23}
\lambda_k(\varphi) \equiv \frac{1}{k} \ln |\mu_k(\varphi)| \qquad
( k=1,2,\ldots) \; ,
\end{equation}
so that the stability condition (19) is equivalent to 
$\lambda_k(\varphi)<0$. Some more information on dynamical systems and 
their stability can be found in Refs. [47--50].

The local stability of an approximation cascade $\{ y_k\}$, that is, the 
local stability of its trajectory $\{ y_k(\varphi)\}$, means the local 
convergence of the approximation sequence $\{ f_k(x)\}$, bijective to 
this trajectory. In other words, if Eq. (19) is correct, then the 
$k$--order approximation is closer to a fixed point, that is more 
accurate, than the initial approximation. And when Eq. (21) is valid, the 
$k$--order approximation is better than the $(k-1)$--order approximation. 
The stability conditions (19) and (21) are sufficient for local 
convergence, but not necessary. So, when such conditions do not hold, 
this does not necessarily mean local divergence, but only tells that 
convergence cannot be guaranteed.

In order to characterize the type of convergence, let us consider the 
variation
$$
\delta y_k(\varphi) \equiv y_k(\varphi +\delta\varphi) - y_k(\varphi)
= \mu_k(\varphi) \delta\varphi \; .
$$
Because of the relation (23) between the local multipliers and Lyapunov 
exponents, we have
\begin{equation}
\label{24}
|\delta y_k(\varphi)| = |\delta\varphi|\exp\{ \lambda_k(\varphi) k\} \; .
\end{equation}
For the stable process, when $\lambda_k(\varphi)<0$, the value $|\delta 
y_k(\varphi)|$ exponentially decreases with $k$. Therefore one tells that 
the calculational procedure is {\it exponentially stable} [51]. This is 
equivalent to saying that the approximation sequence $\{ f_k(x)\}$ 
exponentially converges, since this sequence is, by construction, bijective
to the approximation cascade trajectory $\{ y_k(\varphi)\}$. Similarly, 
we could consider the variation
$$
\delta y_k^*(\varphi) \equiv y_k^*(\varphi+\delta\varphi) -
y_k^*(\varphi) = \frac{\delta y_k^*(\varphi)}{\delta y_k(\varphi)}\;
\mu_k(\varphi)\delta\varphi \; ,
$$
which results in the expression
\begin{equation}
\label{25}
|\delta y_k^*(\varphi)| \leq C\; |\delta\varphi|\;
\exp\{ \lambda_k(\varphi) k \} \; ,
\end{equation}
provided that $\left | \delta y_k^*(\varphi)/\delta y_k(\varphi)\right |
\leq C$. The latter inequality holds true for the stable process, when 
$\lambda_k(\varphi)<0$, then $v_k(\varphi)\rightarrow 0$, as 
$k\rightarrow\infty$, and $y_k(\varphi)\rightarrow y_k^*(\varphi)$, so 
that $|\delta y_k^*(\varphi)/\delta y_k(\varphi)|\rightarrow 1$.

When the values of the sought function $f(x)$ can be found numerically, 
the accuracy of our calculational procedure can be characterized by the 
relative errors
$$
\varepsilon_k(x) \equiv \frac{f_k(x) - f(x)}{|f(x)|}\times 100\% \; ,
\qquad \varepsilon_k^*(x) \equiv \frac{f_k^*(x)-f(x)}{|f(x)|}
\times 100\% \; .
$$
Analysing these, we can judge whether the accuracy of the procedure 
becomes higher with increasing approximation order $k$. But if $f(x)$ is 
not known, how could we decide about the validity of approximations? In 
such a case, we have to be guided by the stability analysis dictating the 
following strategy: (i) When both stability conditions (19) and (21) hold 
true, then we can be certain that we are approaching the correct answer. 
In this case, the minimal time $\tau$ in the evolution integral (17) 
corresponds to the minimal number of steps, that is to $\tau=1$, 
necessary for reaching $f_{k+1}^*$ starting from $f_k$. (ii) When 
condition (19) holds but (21) does not, this means that the $k$--order 
approximation is certainly better than the initial approximation but may 
be worse than the preceeding $(k-1)$--order approximation. In this 
situation, the accuracy of the sought approximation can be improved by 
employing the middle--point method, well known in numerical analysis 
[51]. The middle point can be defined in two ways, explicit and implicit. 
For the evolution integral (17), the explicit definition of the middle 
point implies that we have to associate $f_{k+1}^*$ with the middle of 
the interval between $k$ and $k+1$, that is, we have to set 
$\tau=\frac{1}{2}$. The implicit definition of the middle point between 
two approximations, say $F_{k+1}$ and $F_k$, is to take the average 
$\frac{1}{2}(F_{k+1}+F_k)$ of these approximations [51]. Using this 
average, instead of $F_{k+1}$, in the cascade velocity (16) yields the 
factor $\frac{1}{2}$ for the latter. This evidently is identical to 
putting $\tau=\frac{1}{2}$ in the evolution integral (17). So, the 
explicit and implicit ways of defining the middle point are
equivalent to each other. (iii) Finally, if both conditions (19) and (21) 
are not valid, we cannot trust to the result of calculations. This can 
happen if the choice of control functions or the initial approximation, 
or both of them are not appropriate. Then we need to change these so that 
to achieve the stability of the procedure.

One more question could be as follows. Assume that we can realize our 
calculational procedure in two ways, say using different initial 
approximations. Then which of these ways should we prefer? The answer to 
this question straightforwardly follows from the stability analysis: That 
of several possible ways is preferable which is the most stable.

Thus, in the method of self--similar approximants, there is a unique 
opportunity to control the validity of calculations not knowing the 
sought function and having a small number of terms. In order to unambiguously
prove all this, we consider in the following sections several examples 
for which numerical results are available. Comparing the latter with our 
predictions, everyone can directly observe that the method does work.

\section{Effective Potentials}

Consider the case of the so--called zero--dimensional $\varphi^4$ theory. 
The effective potential, or generating functional, has the structure
\begin{equation}
\label{26}
f(g) = - \ln Z(g) \; ,  \qquad
Z(g) = \frac{1}{\sqrt{\pi}} \int \exp\left\{ - S(g,\varphi)\right\} 
d\varphi \; ,
\end{equation}
with the action
\begin{equation}
\label{27}
S(g,\varphi) = \varphi^2 + g\varphi^4 \; ,
\end{equation}
where the field $\varphi$ is a real variable, $\varphi\in{\bf R}$, and 
the coupling parameter $g\in [0,\infty)$. Expanding $f(g)$ in powers of 
$g$ yields, as is well known, divergent series.

To introduce control functions, we rewrite the action (27) as
$$
S(g,\varphi) = S_0(\varphi,u) + \Delta S(g,\varphi,u) \; ,
$$
\begin{equation}
\label{28}
S_0(\varphi,u) = u^2\varphi^2 \; , \qquad
\Delta S(g,\varphi,u) = ( 1 - u^2)\varphi^2 + g\varphi^4 \; .
\end{equation}
When $\Delta S=0$, we have, in the place of Eq. (26),
\begin{equation}
\label{29}
F_0(g,u) = \ln u \; .
\end{equation}
Invoking the expansion in powers of $\Delta S$, we find
\begin{equation}
\label{30}
F_k(g,u) = F_{k-1}(g,u) + \sum_{p=0}^k c_{kp}\; \alpha^{k-p}\beta^p \; ,
\end{equation}
with the notation
\begin{equation}
\label{31}
\alpha = \alpha(u) \equiv 1 - \frac{1}{u^2} \; , \qquad
\beta = \beta(u) \equiv \frac{3g}{u^4}
\end{equation}
and the coefficients
$$
c_{10} = - \frac{1}{2}\; , \qquad c_{11} = \frac{1}{4} \; , \qquad
c_{20} = - \frac{1}{4} \; , \qquad c_{21} = \frac{1}{2} \; , \qquad 
c_{22} = -\frac{1}{3} \; ,
$$
$$
c_{30} = -\frac{1}{6} \; , \qquad c_{31} = \frac{3}{4} \; , \qquad 
c_{32} = -\frac{4}{3} \; , \qquad c_{33} = \frac{11}{12} \; ,
$$
$$ c_{40} = -\frac{1}{8} \; , \quad c_{41} = 1 \; , \quad 
c_{42} = -\frac{10}{3}\; , \quad c_{43} =\frac{11}{2}\; , \quad
c_{44} = -\frac{34}{9} \; ,
$$
and so on. As is seen, expression (30) is not a series in powers of $g$ 
but a bipolynomial with respect to the variables $\alpha$ and $\beta$ 
defined in Eq. (31).

Control functions $u_k(g)$ are given by the fixed--point condition (15). 
This means that for odd $k=2n+1$, we use the equation $\partial 
F_k/\partial u_k=0$, while for even $k=2n$, for which the latter equation 
has no solution, we put $u_k=u_{k-1}$. This procedure gives
\begin{equation}
\label{32}
u_k(g) =\left [\frac{1}{2}\left ( 1 + \sqrt{ 1 + 12 s_k g}\right )
\right ]^{1/2} \; ,
\end{equation}
where $s_1 = s_2 = 1,\; s_3 = s_4 = 2.239674\; ,\ldots$.
For large $k\gg 1$, one has $s_k\sim k^\varepsilon$, with $1\leq 
\varepsilon\leq 2$. Substituting $u_k(g)$ into Eq. (31), we have the relation
\begin{equation}
\label{33}
\alpha(u_k(g)) \equiv \alpha_k(g) = s_k\beta(u_k(g)) \; ,
\end{equation}
in which
\begin{equation}
\label{34}
\alpha_k(g) = 1 + \frac{1 -\sqrt{1 + 12s_k g}}{6s_k g} \; .
\end{equation}

To construct the approximation cascade, we directly follow the procedure 
of Sec. 2. The sole unimportant difference is that, instead of the 
variable $x$ there, we have here the variable $g$. The equation (4) for 
the expansion function now reads $\ln u_k(g_k(\varphi)) = \varphi$, which
yields
\begin{equation}
\label{35}
g_k(\varphi) = \frac{e^{2\varphi}}{3 s_k}\left ( e^{2\varphi} - 1
\right ) \; .
\end{equation}
The transformation (5) takes the form
\begin{equation}
\label{36}
y_k(\varphi) = \varphi + \sum_{p=1}^k A_{kp} \alpha^p(\varphi) \; ,
\end{equation}
where the coefficients are
\begin{equation}
\label{37}
A_{kp} = \sum_{m=0}^p \frac{c_{pm}}{s_k^m} \qquad ( p \leq k)
\end{equation}
and the notation
\begin{equation}
\label{38}
\alpha(\varphi) \equiv \alpha_k(g_k(\varphi)) = 1 - e^{-2\varphi}
\end{equation}
is used. All coefficients (37) are negative, e.g.
$$
A_{11} = -\frac{1}{4} \; , \qquad A_{21} = A_{11} \; , \qquad
A_{22} = -\frac{1}{12} \; ,
$$
$$
A_{31} = - 0.388377 \; , \qquad A_{32} = - 0.093205 \; , \qquad
A_{33} = - 0.016011 \; ,
$$
$$
A_{41}=A_{31}\; , \qquad A_{42} = A_{32}\; , \qquad A_{43}= A_{33}\; ,
\qquad A_{44} = -0.003606 \; .
$$
The cascade velocity (16) is
\begin{equation}
\label{39}
v_k(\varphi) = B_k\alpha^{k+1}(\varphi) \; , \qquad
B_k \equiv A_{k+1,k+1} \; .
\end{equation}
Invoking the notation
\begin{equation}
\label{40}
f_k^*(g) \equiv \ln\sqrt{1 + z_k^*(g)} \; , \qquad
f_k(g) \equiv \ln\sqrt{1 + z_k(g)} \; ,
\end{equation}
we can write the evolution integral (17) as
\begin{equation}
\label{41}
\int_{z_k}^{z_{k+1}^*} \frac{(1+z)^k}{z^{k+1}}\; dz = 2B_k\tau \; ,
\end{equation}
where $z_k = z_k(g)$ and $z_k^* = z_k^*(g)$. Integrating Eq. (41), we 
obtain the equation
\begin{equation}
\label{42}
z_{k+1}^* = z_k\exp \left\{ P\left (\frac{1}{z_{k+1}^*}\right ) -
P \left ( \frac{1}{z_k}\right ) + 2B_k\tau\right\} \; ,
\end{equation}
defining $z_{k+1}^*(g)$, where
$$
P(x) \equiv \sum_{p=0}^{k-1} C_k^p \; \frac{x^{k-p}}{k-p} \; ,
\qquad C_k^p \equiv \frac{k!}{(k-p)!\; p!} \; .
$$

To check the stability of the approximation cascade $\{ y_k\}$, we 
calculate the local multipliers (18). This gives
\begin{equation}
\label{43}
\mu_k(\varphi) = 1 + 2 [ 1 -\alpha(\varphi) ] \sum_{p=1}^k p\; A_{kp}\;
\alpha^{p-1}(\varphi) \; .
\end{equation}
It is interesting to note the relation between the subsequent multipliers,
\begin{equation}
\label{44}
\mu_{k+1}(\varphi) = \mu_k(\varphi) +\Delta_k(\varphi) \; ,
\end{equation}
where
\begin{equation}
\label{45}
\Delta_k(\varphi) \equiv 2(k+1)\; B_k\; [ 1 - \alpha(\varphi) ]\;
\alpha^k(\varphi) \; .
\end{equation}
From formulas (35) and (38), it follows that the domain of $g\in[0,\infty)$
corresponds to $\varphi\in[0,\infty)$ and $\alpha(\varphi)\in[0,1]$; that 
is, the effective expansion parameter $\alpha(\varphi)$ in Eqs. (36) and 
(43) is always less than $1$ for arbitrary $g\in[0,\infty)$. Also, for 
all $\varphi\in[0,\infty)$ the stability conditions (19) and (21) are 
valid. Therefore, we put $\tau=1$ in Eq. (42).

It is convenient to characterize the accuracy of the approximants (40) by 
the corresponding maximal errors $\varepsilon_k^*\equiv\sup_g
|\varepsilon_k^*(g)|,\;\varepsilon_k \equiv \sup_g | \varepsilon_k(g)|$.
Comparing these approximants with direct numerical calculations for the 
function (26), we have
$\varepsilon_1 = 7\%,\;\varepsilon_2 = 4\%,\;\varepsilon_3 = 0.2\%,\;
\varepsilon_4 = 0.2\% \ldots$, and
$\varepsilon_2^* = 3\%,\; \varepsilon_3^* = 2\%,\;\varepsilon_4^* = 0.1\%
\ldots$, which demonstrates good local convergence of the sequence
$\{ f_k^*(g)\}$ of the self--similar approximants $f_k^*(g)$.

\section{Anharmonic Oscillators}

In $0+1$ dimensions the $\varphi^4$ theory reduces to the familiar 
anharmonic oscillator problem. Therefore, it is instructive to illustrate 
the approach for the eigenvalue problem of quantum mechanics. Let us 
consider the Hamiltonian 
\begin{equation}
\label{46}
H(x) = - \frac{1}{2m} \frac{d^2}{dx^2} + A x^\nu \; ,
\end{equation}
in which $m,A,\nu>0$, and $x\in{\bf R}$. This Hamiltonian, by scaling its 
variable, can be presented in the dimensionless form
\begin{equation}
\label{47}
H = -\frac{1}{2}\frac{d^2}{dx^2} + g x^\nu \; .
\end{equation}
To return from Eq. (47) to Eq. (46), one has to make the substitution
\begin{equation}
\label{48}
H \rightarrow \frac{H(x)}{\omega} \; , \qquad 
x\rightarrow \sqrt{m\omega}\; x \; , \qquad g\rightarrow 1 \; , \qquad
\omega^{2 + \nu} \equiv \frac {A^2}{m^\nu} \; .
\end{equation}
So, we need to find the eigenvalues $e(n,g)$, where $n=0,1,2,\ldots$, of 
the Hamiltonian (47), and then, according to Eq. (48), we have to put 
$g=1$ obtaining the sought spectrum $e(n)\equiv e(n,1)$. The latter can 
be compared with the direct numerical calculation of the Schr\"odinger 
equation, as well as with the quasiclassical approximation [52] which gives
\begin{equation}
\label{49}
e_{WKB}(n) =\left [ \sqrt{\frac{\pi}{2}}\left ( n + \frac{1}{2}\right )
\left ( 1 +\frac{\nu}{2}\right ) \Gamma\left (\frac{2+\nu}{2\nu}\right )
{\Large /} \Gamma\left(\frac{1}{\nu}\right ) \right ]^{2\nu/(2+\nu)} \; ,
\end{equation}
where $\Gamma(\cdot)$ is the Gamma--function.

For $\nu=2$, the Hamiltonian (47) is that of harmonic oscillator. This 
suggests to take for the initial approximation the harmonic oscillator
\begin{equation}
\label{50}
H_0 = - \frac{1}{2}\frac{d^2}{dx^2} + \frac{u^2}{2} \; x^2 \; ,
\end{equation}
in which $u$ is a trial parameter that will generate later control 
functions $u_k(g)$. The eigenvalues and eigenfunctions of $H_0$ are,
respectively,
$$
E_0(n,g,u) =\left ( n +\frac{1}{2}\right ) u \; , \qquad
\psi_n(x) =\frac{(u/\pi)^{1/4}}{(2^n n!)^{1/2}}\;\exp\left ( -\frac{u}{2}
\; x^2  \right ) H_n\left (\sqrt{u}\; x\right ) \; ,
$$
where $n=0,1,2,\ldots$ and $H_n(\cdot)$ is a Hermit polynomial.
By means of the Rayleigh--Schr\"odinger perturbation theory, with respect 
to the perturbation $H-H_0$, one can find the sequence $\{ E_k(n,g,u)\}$ 
of the $k$--order approximations for the eigenvalues of the Hamiltonian 
(47). For what follows, it is convenient to introduce the notation
\begin{equation}
\label{51}
E_k(n,g,u) \equiv \left ( n +\frac{1}{2}\right ) F_k(n,g,u)
\end{equation}
and to consider the sequence $\{ F_k(n,g,u)\}$, with $k=0,1,2,\ldots$ 
Defining control functions $u_k(n,g)$ from the fixed--point equation (15) 
and substituting them into Eq. (51), we get
\begin{equation}
\label{52}
e_k(n) = \left ( n +\frac{1}{2}\right ) \lim_{g\rightarrow 1}
F_k(n,g,u_k(n,g)) \; .
\end{equation}
And from the evolution integral (17), we find the self--similar 
approximant $e_k^*(n)$.

\vskip 3mm

{\bf Quartic oscillator}. Let us realize the above program for $\nu=4$ in
the Hamiltonian (46). For the perturbative terms $F_k$, defined in Eq. (51),
limiting ourselves by the second--order approximation, we have
$$
F_0(n,g,u) = u \; , \qquad
F_1(n,g,u) = F_0(n,g,u) - \frac{u}{2} + \frac{3g}{2u^2}\; \gamma_n \; ,
$$
\begin{equation}
\label{53}
F_2(n,g,u) = F_1(n,g,u) - \frac{u}{8} + \frac{3g}{2u^2}\;\gamma_n -
\frac{5g^2}{2u^5}\; \alpha_n \; ,
\end{equation}
where
$$
\gamma_n \equiv \frac{n^2+n+\frac{1}{2}}{n+\frac{1}{2}} = n +\frac{1}{2}
+ \frac{1}{4\left( n +\frac{1}{2}\right )} \; , \qquad
\alpha_n \equiv 1 + \frac{27}{10}\left ( n +\frac{1}{2}\right )\gamma_n
- \left ( n +\frac{1}{2}\right )^2 \; .
$$
For the control function we obtain $u_1(n,g)=(6\gamma_n\; g)^{1/3}$.
Equation (4) now reads $u_k(n,g_k(n,\varphi))=\varphi$, which results in
the coupling function $g_1(n,\varphi) = \varphi^3/6\gamma_n$.
The endomorphism (5) is written as $y_k(n,\varphi)=\mu_k(n)\varphi$,
where
\begin{equation}
\label{54}
\mu_1(n) = \frac{3}{4}\; , \qquad \mu_2(n) = \frac{3}{4} +\Delta (n)\; ,
\qquad
\Delta(n) \equiv \frac{1}{8} - \frac{5\alpha_n}{72\gamma_n^2} \; .
\end{equation}
The cascade velocity (16) becomes $v_1(n,\varphi) =\Delta(n)\varphi$.
Formula (52) gives
\begin{equation}
\label{55}
e_1(n) =\frac{3}{4}\left ( n +\frac{1}{2}\right ) 
\left ( 6\gamma_n\right )^{1/3} \; , \qquad
e_2(n) =\left [ 1 +\frac{4}{3}\Delta(n) \right ] e_1(n) \; .
\end{equation}
And from the evolution integral (17), we obtain
\begin{equation}
\label{56}
e_2^*(n) = e_1(n)\exp\left\{ \Delta(n)\tau\right\} \; .
\end{equation}

To decide what effective time $\tau$ to choose in the self--similar 
approximant (56), we have to analize the multipliers (18) and (20). For 
the local multipliers (18), we get $\mu_k(n,\varphi) = \mu_k(n)$,
with $\mu_k(n)$ given in Eq. (54). When varying $n=0,1,2,\ldots\infty$,
we have
$$
\frac{35}{48}\leq\mu_2(n)\leq \frac{109}{144} \; , \qquad
-\frac{1}{48} \leq \Delta(n) \leq \frac{1}{144} \; .
$$
Defining the maximal multipliers
\begin{equation}
\label{57}
\mu_k\equiv\sup_n\left |\mu_k(n)\right | \; , \qquad
\overline\mu_k \equiv \sup_n\left |\overline\mu_k(n)\right | \; ,
\end{equation}
we obtain $\mu_1 =0.75,\;\mu_2 = 0.756944,\;\overline\mu_2 = 1.009259$.
It is the maximal multipliers (57) that are to be considered in the 
stability analysis. This is because the sequence $\{ F_k\}$, as in Eq. 
(53), has been derived employing the Raleigh--Schr\"odinger perturbation 
theory, which involves the summation over quantum numbers. Therefore, the 
terms $F_k(n,g,u)$, and consequently the energies (51) and (52), with 
different quantum numbers $n$, cannot be treated as completely independent.

Since $\mu_k<1$, the procedure is locally stable; but $\overline\mu_2$ is 
slightly larger than $1$, so that there is a week ultralocal instability. 
Thus, we need to take the middle point $\tau=\frac{1}{2}$; although, 
because the ultralocal instability is so weak, we could expect that the 
results should not essentially differ between the cases 
$\tau=\frac{1}{2}$ and $\tau=1$.

To check the prediction of the stability analysis, we may compare the 
energy levels (55) and (56) with the results of numerical solution of the 
corresponding Schr\"odinger equation [53]. Varying the quantum numbers 
$n=0,1,2,\ldots$, we define the maximal errors
\begin{equation}
\label{58}
\varepsilon_{WKB} \equiv \sup_n|\varepsilon_{WKB}(n)|\; , \qquad
\varepsilon_k \equiv \sup_n|\varepsilon_k(n)| \; , \qquad
\varepsilon_k^* \equiv \sup_n|\varepsilon_k^*(n)| \; .
\end{equation}
For the latter we find $\varepsilon_{WKB}=1.5\%,\;\varepsilon_1=2\%,\;
\varepsilon_2=0.8\%,\;\varepsilon_2^*=0.4\%$.
The error $\varepsilon_2^*$ does not change for the alternative choices of
$\tau=1$ and $\tau=\frac{1}{2}$.

\vskip 3mm

{\bf Sextic oscillator}. For the power $\nu=6$ in the Hamiltonian (46),
we have the sequence
$$
F_0(n,g,u) = u\; , \qquad
F_1(n,g,u) = F_0(n,g,u) - \frac{u}{2} + \frac{5g}{2u^3}\; \kappa_n \; ,
$$
\begin{equation}
\label{59}
F_2(n,g,u) = F_1(n,g,u) - \frac{u}{8} + \frac{15g}{4u^3}\;\kappa_n -
\frac{g^2}{32u^7}\; \beta_n \; ,
\end{equation}
which is defined in Eq. (51), and where
$$
\kappa_n \equiv n^2 + n +\frac{3}{2} \; , \qquad
\beta_n \equiv 786n^4 + 1572 n^3 + 5324 n^2 + 4538 n + 3495 \; .
$$
Then we follow the same steps as for the quartic oscillator. We find the 
control function $u_1(n,g) = (15\kappa_n\; g)^{1/4}$ and the coupling
function $g_1(n,\varphi)=\varphi^4/15\kappa_n$. Constructing the
approximation cascade, we come to the same endomorphism $y_k(n,\varphi)=
\mu_k(n)\varphi$ but with
\begin{equation}
\label{60}
\mu_1(n) = \frac{2}{3} \; , \qquad \mu_2 = \frac{2}{3} +\Delta(n) \; ,
\qquad
\Delta(n) \equiv \frac{1}{8} - \frac{\beta_n}{7200\kappa_n^2} \; .
\end{equation}
From Eq. (52), we get
\begin{equation}
\label{61}
e_1(n) = \frac{2}{3} \left ( n + \frac{1}{2}\right ) \left ( 
15\kappa_n\right )^{1/4} \; , \qquad
e_2(n) = \left [ 1 +\frac{3}{2}\Delta(n) \right ] e_1(n) \; .
\end{equation}
The integral (17) gives the same expression (56), but with $\Delta(n)$
defined in Eq. (60).

For the local multipliers (18), we have the same form as in the previous
case, with $\mu_k(n)$ given in Eq. (60). Varying the quantum numbers 
$n=0,1,2,\ldots$, we find
$$
\frac{311}{540} \leq \mu_2(n) \leq \frac{273}{400} \; , \qquad
-\frac{19}{1200} \leq \Delta(n) \leq \frac{49}{540} \; .
$$
The maximal multipliers (57) are $\mu_1= 0.6667,\;\mu_2=0.6825,\;
\overline\mu_2=1.023699$. Again we see that the procedure is locally stable
since $\mu_k<1$, but with a weak ultralocal instability because of
$\overline\mu_2>1$. Therefore, we again have to take in Eq. (58) the middle
point $\tau=\frac{1}{2}$. For the maximal errors (58), with respect to
numerical calculations [54], we find $\varepsilon_{WKB} =4\%,\;
\varepsilon_1=7\%,\;\varepsilon_2=8\%,\;\varepsilon_2^*=2\%$.
Taking $\tau=1$ slightly increase the error to $3\%$. Note that in all 
cases the maximal error occurs at the ground--state level.

\vskip 3mm

{\bf Octic oscillator}. With the power $\nu=8$ in the Hamiltonian (46), we
find the sequence of perturbative terms
$$
F_0(n,g,u) = u\; , \qquad
F_1(n,g,u) = F_0(n,g,u) - \frac{u}{2} + \frac{35g}{8u^4}\;\sigma_n \; ,
$$
\begin{equation}
\label{62}
F_2(n,g,u) = F_1(n,g,u) - \frac{u}{8} + \frac{35g}{4u^4}\;\sigma_n -
\frac{g^2}{32u^9}\; \delta_n \; ,
\end{equation}
where
$$
\sigma_n \equiv \frac{n^4 + 2n^3 + 5n^2 + 4n + 3/2}{n+ 1/2} \; ,
$$
$$
\delta_n \equiv 3985 n^6 + 11955 n^5 + 74904 n^4 + 129883 n^3 +
277901 n^2 + 214952 n + 135030\; .
$$ 
The control function is $u_1(n,g)=(35\sigma_n\; g)^{1/5}$
and the coupling function becomes $g_1(n,\varphi)=\varphi^5/35\sigma_n$.
The approximation cascade is described by the same endomorphism with
\begin{equation}
\label{63}
\mu_1(n) = \frac{5}{8} \; , \qquad \mu_2 = \frac{5}{8} +\Delta(n) \; ,
\qquad
\Delta(n) \equiv \frac{1}{8} - \frac{\delta_n}{39200\sigma_n^2} \; .
\end{equation}
For varying $n=0,1,2,\ldots$, one has
$$
\frac{4319}{11760} \leq \mu_2(n) \leq \frac{5083}{7840} \; , \qquad
-\frac{3031}{11760} \leq \Delta(n) \leq \frac{183}{7840} \; .
$$
The energies (52) are
\begin{equation}
\label{64}
e_1(n) = \frac{5}{8} \left ( n + \frac{1}{2}\right ) \left ( 
35\sigma_n\right )^{1/5} \; , \qquad
e_2(n) = \left [ 1 +\frac{8}{5}\Delta(n) \right ] e_1(n) \; .
\end{equation}
And the self--similar approximant $e_2^*(n)$ again has the same form (56).

The local multipliers are given by $\mu_k(n)$. For the maximal multipliers 
(57) we find $\mu_1=0.625,\;\mu_2=0.648342,\;\overline\mu_2=1.037347$.
This shows that the procedure is locally stable but is not ultralocally 
stable. So, the self--similar approximant (56) has to be defined at the 
middle point $\tau=\frac{1}{2}$.

For the maximal errors (58), with respect to numerical calculations [54], 
we obtain $\varepsilon_{WKB}=7\%,\;\varepsilon_1=13\%,\;
\varepsilon_2=34\%,\;\varepsilon_2^*=3\%$. 
The accuracy of the self--similar approximant (56) for $\tau=\frac{1}{2}$
is essentially higher than that for $\tau=1$, for which case the error 
reaches $13\%$.

\vskip 3mm

{\bf Multidimensional oscillator}. In the previous three cases we have
shown that the method works well for different one--dimensional anharmonic
oscillators. Now we want to demonstrate that it is equally applicable for
anharmonic oscillators of arbitrary dimensionality. For this purpose, we
consider the spherically symmetric quartic oscillator in $D$--dimensional
real space. The corresponding radial Hamiltonian reads
\begin{equation}
\label{65}
H(r) = -\frac{1}{2m} \frac{d^2}{dr^2} + \frac{(2l+D-3)(2l+D-1)}{8mr^2} +
Ar^4 \; ,
\end{equation}
where $m,A>0;\; l=0,1,2,\ldots;\; r\geq 0$, and $D=1,2,3,\ldots$ is a 
real--space dimensionality. This Hamiltonian, by the appropriate scaling, 
can be reduced to the form
\begin{equation}
\label{66}
H = -\frac{1}{2} \frac{d^2}{dr^2} + \frac{(2l+D-3)(2l+D-1)}{8r^2} 
+gr^4 \; .
\end{equation}
One can return from Eq. (66) back to Eq. (65) by means of the substitution
$$
H\rightarrow \frac{H(r)}{\omega}\; , \qquad r\rightarrow 
\sqrt{m\omega}\; r\; , \qquad g\rightarrow 1\; , \qquad 
\omega\equiv \left (\frac{A}{m^2}\right )^{1/3} \; .
$$

We start perturbation theory with the harmonic $D$--dimensional oscillator
defined by the Hamiltonian
$$
H_0 = -\frac{1}{2}\frac{d^2}{dr^2} + \frac{(2l+D-3)(2l+D-1)}{8r^2}
+\frac{u^2}{2}\; r^2 \; ,
$$
in which $u$ is a trial parameter. The corresponding eigenvalues are
$$
E_0(n,l,u) = \left ( 2n + l + \frac{D}{2}\right ) u \; ,
$$
where $n=0,1,2,\ldots$, and the eigenfunctions are
$$
\chi_{nl}(r) = \left [ \frac{2n!u^{l+D/2}}{\Gamma(n+l+D/2)}\right ]^{1/2}
r^{(2l+D-1)/2}\exp\left ( -\frac{u}{2}r^2\right ) 
L_n^{(2l+D-2)/2}(ur^2) \; ,
$$ where $L_n^l(r)$ is the associate Laguerre polynomial. The latter can 
be presented in several forms, so we write down below that one we use in 
what follows:
$$
L_n^l(r) = \frac{1}{n!}\; e^r\; r^{-l}\; \frac{d^n}{dr^n}\left (
e^{-r} r^{n+l}\right ) = 
\sum_{m=0}^n \frac{\Gamma(n+l+1)(-r)^m}{\Gamma(m+l+1)m!(n-m)!}\; .
$$

For the terms of perturbation theory, we use the notation
\begin{equation}
\label{67}
E_k(n,l,g,u) \equiv \left ( 2n + l +\frac{D}{2}\right ) F_k(n,l,g,u)\; .
\end{equation}
Then, in the second order we have
$$
F_0(n,l,g,u) = u\; , \qquad
F_1(n,l,g,u) = F_0(n,l,g,u) - \frac{u}{2} + \frac{3g}{2u^2}\;\gamma_{nl} \; ,
$$
\begin{equation}
\label{68}
F_2(n,l,g,u) = F_1(n,l,g,u) - \frac{u}{8} + \frac{3g}{2u^2}\;\gamma_{nl} -
\frac{5g^2}{2u^5}\; \alpha_{nl} \; ,
\end{equation}
where
$$
\gamma_{nl} \equiv 2n +l + \frac{D}{2} - 
\frac{(l+D/2)(l-2+D/2)}{3(2n+l+D/2)} \; ,
$$
$$
\alpha_{nl} \equiv 1 + \frac{27}{10} \left ( 2n + l +\frac{D}{2}\right )
\gamma_{nl} - \left ( 2n +l +\frac{D}{2}\right )^2 \; .
$$
The one--dimensional case can be recovered with the change 
$D\rightarrow 1$, $2n\rightarrow n$, $l\rightarrow 0$.

All steps of the procedure are the same as earlier, so we do not go into 
much details. For the control functions we find
$u_1(n,l,g) = (6\gamma_{nl} g)^{1/3}$, and for the coupling function we get
$g_1(n,l,\varphi)=\varphi^3/6\gamma_{nl}$.

The approximation cascade is characterized by the endomorphism
$y_k(n,l,\varphi)=\mu_k(n,l)\varphi$,
in which
$$
\mu_1(n,l) = \frac{3}{4} \; , \qquad \mu_2(n,l) = \frac{3}{4} +
\Delta(n,l) \; , \qquad
\Delta(n,l) \equiv \frac{1}{8} - \frac{5\alpha_{nl}}{72\gamma_{nl}^2} \; ,
$$
and the cascade velocity is $v_1(n,l,\varphi)=\Delta(n,l)\varphi$.

Similarly to Eq. (52), we define
$$
e_k(n,l) = \left ( 2n +l +\frac{D}{2}\right ) \lim_{g\rightarrow 1}
F_k(n,l,g,u_k(n,l,g)) \; .
$$
This yields
\begin{equation}
\label{69}
e_1(n,l) =\frac{3}{4}\left ( 2n +l +\frac{D}{2}\right ) 
(6\gamma_{nl})^{1/3} \; , \qquad
e_2(n,l) =\left [ 1 +\frac{4}{3}\Delta(n,l)\right ] e_1(n,l) \; .
\end{equation}
And the evolution integral (17) gives us the self--similar approximant
\begin{equation}
\label{70}
e_2^*(n,l) = e_1(n,l) \exp\left\{ \Delta(n,l)\tau\right\} \; .
\end{equation}

The stability analysis here is practically the same as for the 
one--dimensional quartic oscillator, with the same conclusion that the 
procedure is locally stable but with a very weak ultralocal instability. 
Consequently, in the approximant (70) one has to take the middle point 
$\tau=\frac{1}{2}$, although the results should not be much different 
from the case $\tau=1$. The accuracy should increase with the increasing 
dimensionality $D$, since increasing $D$ supresses ultralocal 
instability. For example, when $D\rightarrow\infty$, then 
$\gamma_{nl}\rightarrow D/3$, $\alpha_{nl}\rightarrow D^2/5$, and
$\Delta(n,l)\rightarrow 0$, so that $\overline\mu_2\rightarrow 1$. These 
conclusions are in perfect agreement with calculations. Thus, the maximal 
error of the self--similar approximant (70) is $\varepsilon_2^*=0.4\%$ 
for $D=1$ and it is $\varepsilon_2^*=0.3\%$ for $D=3$.

\section{Arbitrary Powers}

In the previous section, we considered several Hamiltonians with 
potentials having even integer powers. However, this restriction is not 
principal, and the method can be applied to the case of potentials with 
arbitrary, noninteger as well as integer, powers. To demonstrate this, 
let us consider a spherically symmetric three--dimensional problem with 
the radial Hamiltonian
\begin{equation}
\label{71}
H(r) = -\frac{1}{2m} \frac{d^2}{dr^2} + \frac{l(l+1)}{2mr^2} +
Ar^\nu \; ,
\end{equation}
in which $m,A>0;\; l=0,1,2,\ldots;\; r\geq 0$; and $\nu>0$ is an arbitrary
positive number. By scaling, it is again convenient to reduce this 
Hamiltonian to the simplified form
\begin{equation}
\label{72}
H = -\frac{1}{2}\frac{d^2}{dr^2} + \frac{l(l+1)}{2r^2} + gr^\nu \; .
\end{equation}
The return from Eq. (72) to (71) is made by means of the substitution
$$
H\rightarrow \frac{H(r)}{\omega}\; , \qquad
r\rightarrow \sqrt{m\omega}\; r\; , \qquad g\rightarrow 1 \; , \qquad
\omega^{2+\nu}\equiv \frac{A^2}{m^\nu} \; .
$$
Hamiltonians of this type have been considered in a number of papers (see 
e.g. [31,55--58] and references therein). The importance of potentials 
with various powers is due to the fact that such potentials model well 
confining forces between heavy quarks and lead to a reasonable description
of the quarkonium spectrum [59]. The quasiclassical approximation gives [59]
the eigenvalues of the Hamiltonian (72) in the form
\begin{equation}
\label{73}
e_{WKB}(n,l) =\left [ \sqrt{\frac{\pi}{2}}\left ( 2n + l + \frac{3}{2}
\right ) \left ( 1 +\frac{\nu}{2}\right ) 
\Gamma\left (\frac{2+\nu}{2\nu}\right ) {\Large /} 
\Gamma\left ( \frac{1}{\nu}\right )\right ]^{2\nu/(2+\nu)}\; .
\end{equation}

Applying our method to the eigenvalue problem for the Hamiltonian (72), 
we start with the harmonic Hamiltonian
$$
H_0 = -\frac{1}{2}\frac{d^2}{dr^2} + \frac{l(l+1)}{2r^2} +
\frac{u^2}{2}r^2 \; ,
$$
whose eigenvalues are
$$
E_0(n,l,u) =\left ( 2n + l +\frac{3}{2}\right ) u \; .
$$
Using the Rayleigh--Schr\"odinger perturbation theory, we can find the 
$k$--order terms
\begin{equation}
\label{74}
E_k(n,l,g,u) \equiv \left ( 2n + l +\frac{3}{2}\right ) F_k(n,l,g,u)\; .
\end{equation}
Then, defining control functions $u_k(n,l,g)$, we come to the 
renormalized eigenvalues
\begin{equation}
\label{75}
e_k(n,l) \equiv \left ( 2n + l +\frac{3}{2}\right ) \lim_{g\rightarrow 1}
F_k(n,l,g,u_k(n,l,g)) \; .
\end{equation}

In the second order, we obtain the sequence
$$
F_0(n,l,g,u) = u \; , \qquad
F_1(n,l,g,u) = F_0(n,l,g,u) - \frac{u}{2} + \frac{g}{u^{\nu/2}}\;
A_{nl}\left (\frac{\nu}{2}\right ) \; ,
$$
\begin{equation}
\label{76}
F_2(n,l,g,u) = F_1(n,l,g,u) - \frac{u}{8} + 
\frac{g}{2u^{\nu/2}}\; B_{nl}\left (\frac{\nu}{2}\right ) -
\frac{g^2}{2u^{1+\nu}}\; C_{nl}\left (\frac{\nu}{2}\right ) \; ,
\end{equation}
with the coefficients
$$
A_{nl}(\nu) \equiv \frac{n!\; I_{nn}^l(\nu)}{(2n+l+3/2)\Gamma(n+l+3/2)} \; ,
$$
$$
B_{nl}(\nu) \equiv \frac{n!}{(2n+l+3/2)\Gamma(n+l+3/2)}
\left [ (n+1) I_{n,n+1}^l(\nu) - \left ( n+l+\frac{1}{2}\right)
I_{n,n-1}^l(\nu)\right ]\; ,
$$
$$
C_{nl}(\nu) \equiv \frac{n!}{(2n+l+3/2)\Gamma(n+l+3/2)}
\sum_{p=0\;(p\neq n)}^\infty
\frac{p!\left [ I_{np}^l(\nu)\right ]^2}{(p-n)\; \Gamma(p+l+3/2)}\; .
$$
Here we use the notation
\begin{equation}
\label{77}
I_{ns}^l(\nu) \equiv \int_0^\infty t^{l+\nu+1/2}\; e^{-t}\; L_n^{l+1/2}(t)\;
L_s^{l+1/2}(t)\; dt \; ,
\end{equation}
where $L_n^l(t)$ is an associate Laguerre polynomial. The way of calculating 
this integral is explained in Appendix A. As is seen from the sequence 
(76), we have there an expression in powers of $g/u^{1+\nu/2}$.

Following the same steps as in the previous section, we find the control and
coupling functions
$$
u_1(n,l,g) =\left [ \nu A_{nl}\left (\frac{\nu}{2}\right ) 
g\right ]^{2/(2+\nu)}\; , \qquad
g_1(n,l,\varphi) = \frac{\varphi^{1+\nu/2}}{\nu A_{nl}(\nu/2)}\; .
$$
The approximation cascade is defined by the endomorphism $y_k(n,l,\varphi)=
\mu_k(n,l)\varphi$ with
$$
\mu_1(n,l) = \frac{2+\nu}{2\nu}\; , \qquad \mu_2(n,l) = \frac{2+\nu}{2\nu}
+ \Delta(n,l) \; ,
$$
\begin{equation}
\label{78}
\Delta(n,l) \equiv \frac{B_{nl}(\nu/2)}{2\nu A_{nl}(\nu/2)} -
\frac{C_{nl}(\nu/2)}{2\nu^2 A_{nl}(\nu/2)} -\frac{1}{8} \; .
\end{equation}
For the values (75), we have
$$
e_1(n,l) =\left ( 2n + l +\frac{3}{2}\right ) \frac{2+\nu}{2\nu} 
\left [ \nu A_{nl}\left (\frac{\nu}{2}\right ) \right ]^{2/(2+\nu)}\; ,
$$
\begin{equation}
\label{79}
e_2(n,l) = \left [ 1 +\frac{2\nu}{2+\nu}\Delta(n,l)\right ] e_1(n,l) \; .
\end{equation}
The evolution integral (17) yields the self--similar approximant
\begin{equation}
\label{80}
e_2^*(n,l) = e_1(n,l) \exp\left\{ \Delta(n,l)\tau\right \} \; .
\end{equation}

The local multipliers (18) are $\mu_k(n,l,\varphi) = \mu_k(n,l)$,
with the right--hand side defined in Eq. (78). The analysis of these 
multipliers shows that the stability properties depend on the power $\nu$ 
of the potential in the Hamiltonian (72). For $\nu\neq 2$, the procedure 
is either locally unstable but then it is ultralocally stable, or the 
procedure is locally stable but ultralocally unstable. In both these 
cases, we have to take in the approximant (80) the middle point 
$\tau=\frac{1}{2}$. The case $\nu=2$ is special. Then all multipliers 
$\mu_k=\overline\mu_k = 1$, which corresponds to the neutral stability. This 
means that we are already at a fixed point of the cascade trajectory, and 
there is no motion any more. Really, our calculations show that for 
$\nu=2$ and any $\tau$, we have for all eigenvalues 
$e_1(n,l)=e_2(n,l)=e_2^*(n,l)$, which coincides with the exact solution 
for the harmonic oscillator. To illustrate what are the values of typical 
errors characterizing the accuracy of different approximations, we 
present in Table I the results of calculations of the ground--state 
energy for various powers $\nu$. We give there the errors $\varepsilon_1$ 
and $\varepsilon_2$ for the renormalized expressions from Eq. (79), the 
error $\varepsilon_2^*$ for the self--similar approximant (80), and also 
the error of the quasiclassical approximation (73). These errors are 
computed by comparing the results of our calculations with accurate 
numerical data [55,57,59] obtained by the direct numerical solution of 
the Schr\"odinger equation. The results for the approximant (80) 
correspond to the middle--point case $\tau=\frac{1}{2}$, which is taken 
according to the stability analysis. For comparison, we also made 
calculations setting $\tau=1$. The latter case leads to the errors that 
are close to $\varepsilon_2^*$ from Table I, when $\nu\leq 6$. However, 
for $\nu>6$, the errors quickly increase becoming, e.g., $-6\%$ for 
$\nu=8$ and $-19\%$ for $\nu=10$. Hence, the middle point choice 
$\tau=\frac{1}{2}$ is especially preferable for large powers of the 
potential.

\section{Initial Approximations}

Now we aim at analysing the following problem. Assume that we can take 
not just one but two or more initial approximations that can be employed 
as starting points for constructing the corresponding approximation 
cascades. Then how to decide which of these initial approximations is 
preferable? We show that the stability analysis again can give us an 
answer to this question.

For the purpose under consideration, it is convenient to take the 
logarithmic potential whose behaviour is such that it can be modelled, to 
some extent, either by the harmonic oscillator or by the Coulomb 
potential. We mean here the three--dimensional spherically symmetric 
problem with the radial Hamiltonian
\begin{equation}
\label{81}
H(r) = -\frac{1}{2m}\frac{d^2}{dr^2} + \frac{l(l+1)}{2mr^2} + 
B\ln\frac{r}{b} \; ,
\end{equation}
in which $m,B,b>0$, and $r\geq 0$. As usual, we transform the Hamiltonian 
(81) to the scaled form
\begin{equation}
\label{82}
H = -\frac{1}{2}\frac{d^2}{dr^2} + \frac{l(l+1)}{2r^2} + 
g\ln r \; .
\end{equation}
The return from Eq. (82) to Eq. (81) is made by means of the substitution
$$
H\rightarrow \frac{H(r)}{\omega} \; , \qquad r\rightarrow \frac{r}{b}\; ,
\qquad g\rightarrow \frac{B}{\omega} \; , \qquad \omega \equiv 
\frac{1}{mb^2}\; .
$$
Note that the logarithmic potential is also often used for describing 
confining forces between quarks [59], and the quasiclassical approximation 
for the spectrum of the Hamiltonian (82) is known to be
\begin{equation}
\label{83}
e_{WKB}(n,l,g) = \frac{g}{2}\ln\left [ \frac{\pi}{2g} \left ( 2n +l +
\frac{3}{2}\right )^2 \right ] \; .
\end{equation}

\vskip 3mm

{\bf Harmonic potential}. Let us choose for the initial approximation the
harmonic--oscillator Hamiltonian. Then for the function
\begin{equation}
\label{84}
F_k(n,l,g,u) \equiv \frac{E_k(n,l,g,u)}{(2n+l+3/2)}
\end{equation}
we derive the sequence of perturbative terms
$$
F_0(n,l,g,u) = u \; ,
$$
\begin{equation}
\label{85}
F_1(n,l,g,u) = F_0(n,l,g,u) - \frac{u}{2} + g \;
\frac{\psi(n+l+3/2)-\ln u}{2(2n+l+3/2)}\;,
\end{equation}
$$
F_2(n,l,g,u) = F_1(n,l,g,u) - \frac{u}{8} + 
\frac{g}{4(2n+l+3/2)} - \frac{g^2}{8u}\; C_{nl}\; , 
$$
where
$$
\psi(x) \equiv \frac{d}{dx}\ln\Gamma(x) \; ,
$$
$$
C_{nl} \equiv \frac{n!}{(2n+l+3/2)\Gamma(n+l+3/2)}
\sum_{p=0\; (p\neq n)}^\infty 
\frac{p!\left ( I_{np}^l\right )^2}{(p-n)\Gamma(p+l+3/2)}\; ,
$$
Here $I_{ns}^l$ denotes the integral
\begin{equation}
\label{86}
I_{ns}^l \equiv \int_0^\infty t^{l+1/2}\; e^{-t}\; \ln t\; L_n^{l+1/2}(t)\;
L_s^{l+1/2}(t)\; dt \; .
\end{equation}
The properties of this integral are described in Appendix B.

Following the standard scheme of our method, we find the control
and coupling functions
$$
u_1(n,l,g) = \frac{g}{2n+l + 3/2}\; , \qquad
g_1(n,l,\varphi) = \left ( 2n + l +\frac{3}{2}\right )\varphi \; .
$$
Then we construct the approximation cascade $\{ y_k\}$, with the 
trajectory points
\begin{equation}
\label{87}
y_1(n,l,\varphi) = \frac{\varphi}{2}\left [ 1 -\ln\varphi +
\psi\left ( n + l +\frac{3}{2}\right )\right ] \; , \qquad
y_2(n,l,\varphi) = y_1(n,l,\varphi) + v_1(n,l,\varphi) \; ,
\end{equation}
and with the cascade velocity $v_1(n,l,\varphi)=\Delta(n,l)\varphi$,
where
$$
\Delta(n,l) \equiv \frac{1}{8} - \frac{1}{8}\left ( 2n +l +\frac{3}{2}
\right )^2 C_{nl} \; , \qquad \Delta(0,0) = -0.018288 \; .
$$
For the renormalized expressions
\begin{equation}
\label{88}
e_k(n,l,g) =\left ( 2n +l +\frac{3}{2}\right ) F_k(n,l,g,u_k(n,l,g))\; ,
\end{equation}
we have
$$
e_1(n,l,g) =\frac{g}{2}\left [ 1 +\ln\frac{2n+l+3/2}{g} + 
\psi\left ( n + l +\frac{3}{2}\right ) \right ] \; ,
$$
\begin{equation}
\label{89}
e_2(n,l,g) = e_1(n,l,g) + \Delta(n,l)\; g \; .
\end{equation}
From the evolution integral (17), we obtain the self--similar approximant
\begin{equation}
\label{90}
e_2^*(n,l,g) = e_1(n,l,g)\exp\left\{ \Delta(n,l)\tau\right\} \; .
\end{equation}
The qualitative behaviour of the energy (90) is as follows. At small 
$g\rightarrow+0$, this energy is positive and tends to zero as $-g\ln g$. 
As $g$ increases, $e_2^*$ increases till the maximum value
$$
e_2^*(n,l,g_{max}) =
\frac{1}{2} \; g_{max}(n,l) \exp\{ \Delta(n,l)\tau \} \; ,
$$ occurring at
\begin{equation}
\label{91}
g_{max}(n,l) =\left ( 2n+l +\frac{3}{2}\right ) \exp\left\{
\psi\left ( n + l +\frac{3}{2}\right ) \right\} \; .
\end{equation}
With the further increase of $g$, function (90) decreases to zero,
$e_2^*(n,l,g_0) = 0$, at the value
\begin{equation}
\label{92}
g_0(n,l) = eg_{max}(n,l) \; ,
\end{equation}
after which $e_2^*$ remains negative. Using notation (92), we may 
present Eq. (90) as
\begin{equation}
\label{93}
e_2^*(n,l,g) = \frac{g}{2} \ln\left ( \frac{g_0}{g}\right ) \exp\left\{
\Delta(n,l)\tau\right \} \; ,
\end{equation}
where $g_0=g_0(n,l)$.

To estimate the values of the characteristic couplings (91) and (92), 
consider the case n=l=0, when
\begin{equation}
\label{94}
g_{max}(0,0) = \frac{3}{8}\; e^{2-C}=1.555746\; , \qquad
g_0(0,0) = \frac{3}{8}\; e^{3-C}=4.228956 \; .
\end{equation}
where $C$ is the Euler constant.

In the opposite case, when either $n$ or $l$ is large, since 
$\psi(x)\simeq \ln x$, as $x\rightarrow\infty$, we have
\begin{eqnarray}
g_{max}(n,l) \simeq \left\{ 
\begin{array}{cc}
2n^2 & (n\rightarrow\infty,\; l<\infty) \\
\label{95}
l^2  & (n<\infty,\; l\rightarrow\infty) \; .
\end{array}\right.
\end{eqnarray}

The local multipliers (18), for the trajectories points (87), are
\begin{equation}
\label{96}
\mu_1(n,l,\varphi) = \frac{1}{2}\psi\left ( n + l +\frac{3}{2}\right ) -
\frac{1}{2}\ln\varphi \; , \qquad
\mu_2(n,l,\varphi) = \mu_1(n,l,\varphi) + \Delta(n,l) \; .
\end{equation}
The images of the multipliers (96), defined in Eq. (22), become
$$
m_1(n,l,g) = \frac{1}{2}\; \psi\left ( n + l + \frac{3}{2}\right ) +
\frac{1}{2}\; \ln\frac{2n+l+3/2}{g} \; ,
$$
\begin{equation}
\label{97}
m_2(n,l,g) = m_1(n,l,g) + \Delta(n,l) \; .
\end{equation}
With definition (91), we may write
\begin{equation}
\label{98}
m_1(n,l,g) = \frac{1}{2}\; \ln\frac{g_{max}(n,l)}{g} \; .
\end{equation}
Varying the quantum numbers $n$ and $l$, one can check that $\Delta(n,l)$ 
is always negative and the ultralocal multiplier
$\overline m_2(n,l,g)\equiv m_2(n,l,g)/m_1(n,l,g)$, which is the image of
the multiplier (20), is less than one for all $n,\; l$, and $g$. That is,
the procedure is ultralocally stable. However, it is locally stable not for
all coupling parameters but only for those satisfying the inequalities
\begin{equation}
\label{99}
\frac{1}{e^2} < \frac{g}{g_{max}} < e^2 \; ,
\end{equation}
where $g_{max}=g_{max}(n,l)$ is given by Eq. (91). For instance, if 
$n=l=0$, then the interval (99) is
\begin{equation}
\label{100}
0.210547 < g < 11.495494\; .
\end{equation}
In the region of stability (100), we may put $\tau=1$. For illustration, we 
present in Table II the results of calculations for $n=0$ and $l=0,1,2,3,4$, 
and for the coupling parameter $g=0.5$. We show the percentage errors 
$\varepsilon_1$ and $\varepsilon_2$ for the renormalized expressions (89), 
the error $\varepsilon_2^*$ of the self--similar approximant (90), and the 
error $\varepsilon_{WKB}$ of the quasiclassical approximation (83). The 
comparison is made with respect to numerical data [55,59].

\vskip 3mm

{\bf Coulomb potential}. Now let us turn to the case when the Hamiltonian
\begin{equation}
\label{101}
H_0 = -\frac{1}{2}\frac{d^2}{dr^2} + \frac{l(l+1)}{2r^2} + 
\frac{u}{r} \; 
\end{equation}
with the Coulomb potential is chosen for the initial approximation. The 
corresponding eigenvalues and eigenfunctions are
$$
E_0(n,l,g,u) = -\frac{u^2}{2(n+l+1)^2}\; ,
$$
$$
\chi_{nl}(r) =\left [\frac{n!\; u}{(n+2l+1)!}\right ]^{1/2}\frac{1}{n+l+1}
\left (\frac{2ur}{n+l+1}\right )^{l+1}\exp\left ( -\frac{ur}{n+l+1}\right )
L_n^{2l+1}\left ( \frac{2ur}{n+l+1}\right ) \; .
$$
In calculating matrix elements, we meet the integral
\begin{equation}
\label{102}
J_{ns}^l \equiv \int_0^\infty t^{2l+2}\; e^{-t}\; \ln t\; L_n^{2l+1}(t)\;
L_s^{2l+1}(t)\; dt \; ,
\end{equation}
whose properties are specified in Appendix C. The whole calculational 
procedure is very similar to the previous subsection, because of which we 
shorten here technical details. This is especially justified by the fact 
that the procedure starting with the initial approximation corresponding 
to the Hamiltonian (101), as will be shown, is less stable than that 
beginning with the harmonic--oscillator Hamiltonian.

The spectrum in the first approximation reads
\begin{equation}
\label{103}
E_1(n,l,g,u) = \frac{u^2}{2(n+l+1)^2} + g\left ( \ln\frac{n+l+1}{2u} +
D_{nl}\right ) \; ,
\end{equation}
where
$$
D_{nl} \equiv \frac{2n+1}{2(n+l+1)} + \psi(n+2l+2) \; .
$$
For the control and coupling functions, we find
$$
u_1(n,l,g) \equiv (n+l+1)\sqrt{g}\; , \qquad
g_1(\varphi) = -2\varphi \; .
$$
Then the point of the approximation cascade, bijective to the 
approximation (103), is
$$
y_1(n,l,\varphi) = -\varphi [ 1 - \ln(-8\varphi) + 2D_{nl} ] \; .
$$
Substituting the control function into spectrum (103) gives
\begin{equation}
\label{104}
e_1(n,l,g) =\frac{g}{2} \left ( 1 +\ln\frac{1}{4g} + 2D_{nl}\right ) \; .
\end{equation}
The self--similar approximant $e_2^*$ acquires the form of Eq. (90).

The overall qualitative behaviour of the spectrum is the same as in the 
previous subsection. For $g\rightarrow 0$, the energy tends to zero as 
$-g\ln g$. With increasing $g$, the energy reaches a maximum at the point
\begin{equation}
\label{105}
g_{max}(n,l) = \frac{1}{4}\exp (2D_{nl}) \; .
\end{equation}
Then the energy diminishes to zero at the value $g_0=g_0(n,l)$ having the 
same relation $g_0=eg_{max}$, as in Eq. (92). For the ground--state 
level, we now have
\begin{equation}
\label{106}
g_{max}(0,0) = \frac{1}{4}\; e^{3-2C} = 1.582925 \; , \qquad
g_0(0,0) = \frac{1}{4}\; e^{4-2C} = 4.302836 \; .
\end{equation}
The energy (104) can be written as
\begin{equation}
\label{107}
e_1(n,l,g) = \frac{g}{2}\; \ln\left (\frac{g_0}{g}\right ) \; ,
\end{equation}
where $g_0=g_0(n,l)$, and $e_2^*$ can be presented as in Eq. (93).

The characteristic values of $g_{max}$, given by Eqs. (105) and (91), 
are very close to each other. This is seen from Eqs. (94) and (106), as 
well as comparing Eq. (95) with the corresponding behaviour of Eq. (105) 
yielding
\begin{eqnarray}
g_{max}(n,l) \simeq \left\{ 
\begin{array}{cc}
(e^2/4)n^2 & (n\rightarrow\infty,\; l < \infty) \\
\nonumber
l^2  & (n < \infty, \; l\rightarrow\infty) \; .
\end{array}\right.
\end{eqnarray}

For the local multiplier (18) and its image (22), we find
\begin{equation}
\label{108}
\mu_1(n,l,\varphi) = \ln(-8\varphi) -2D_{nl} \; , \qquad
m_1(n,l,g) = \ln\frac{g}{g_{max}(n,l)} \; .
\end{equation}
The stability condition $|\mu_1(n,l,\varphi)|<1,\;|m_1(n,l,\varphi)|<1$ is
satisfied in the region
\begin{equation}
\label{109}
\frac{1}{e} < \frac{g}{g_{max}} < e \; ,
\end{equation}
where $g_{max}=g_{max}(n,l)$ is defined in Eq. (105). For the ground 
state, the inequalities (109) give
\begin{equation}
\label{110}
0.582326 < g < 4.302836 \; .
\end{equation}

Comparing the stability properties of the calculational procedure for the 
considered cases, when either the harmonic oscillator or the Coulomb 
potential are used for the initial approximation, we come to the 
following conclusion. The region of stability, with respect to the 
coupling parameter $g$, is three times narrower in the latter case than 
in the former. Also, the values of the local multipliers in the latter 
case are about twice as large as for the former case. Consequently, the 
calculational procedure starting with the harmonic oscillator is more 
stable than that beginning with the Coulomb potential. This conclusion is 
in agreement with calculations showing that the first, more stable, case 
is also more accurate than the second, less stable; the error of the 
latter case being almost twice larger.

\section{Conclusion}

The self--similar perturbation theory makes it possible to obtain quite 
accurate approximations for those complicated physical problems for which 
only a few terms of a perturbative algorithm are available. The term 
"perturbation theory" here has to be understood in the general mathematical 
sense as a regular procedure yielding a set of subsequent approximations. 
This approach does not require any small parameters of the physical problem 
considered. For instance, the coupling parameters can be arbitrary strong. 
Convergence is achieved by introducing control functions, by means of which 
the sequence of approximations is reorganized to become convergent for any 
given parameters. 

The approach is based on the property of self--similarity between 
approximations. This property, for the family of endomorphisms representing 
the given approximation sequence, is a necessary condition for the motion in 
the vicinity of a fixed point. For the approximation sequence itself, this 
means a necessary condition for the fastest convergence [42].

Let us stress the difference between the {\it self--similarity of 
approximations} and the functional self--similarity which is the basis of 
any renormalization--group approach [46,60]. In the latter case, one looks 
for a symmetry of the considered function $f(x)$ with respect to the scaling 
of the variable $x$, so that to find a relation $f(\lambda x)=B(\lambda,f(x))$.
From such a relation, it is easy to get the renormalization group equation 
$x\partial f(x)/\partial x=\beta(f(x))$, where $\beta(f)\equiv
\lim_{\lambda\rightarrow 1}\partial B(\lambda,f)/\partial\lambda$ is called
the renormalization group function. When the above relation is exact, one 
speaks of an exact renormalization group. However, in many cases such 
relations can be derived only approximately. As is clear from comparing 
this standard renormalization group technique with our approach, we do 
not consider the scaling of variables, but instead we are analysing a 
kind of {\it scaling of approximation numbers}. This what makes our 
approach principally different from any variant of the renormalization 
group techniques.

The property of the self--similarity of approximations permits us to 
reformulate perturbation theory to the language of dynamical theory and 
optimal control theory. Such a reformulation allows us to resort to 
powerful methods of these theories in order to give a logical foundation 
for the whole approach. Thus, it immediately becomes clear that control 
functions are to be defined by the fixed--point equations. Then, 
considering the motion near a fixed point, we find nontrivial corrections 
essentially improving the accuracy of approximations. And, what is 
probably the most important, we can use the stability analysis in order 
to check the stability of the procedure and, respectively, the 
convergence of the approximation sequence. This is based on the analysis 
of the local multipliers that are directly related to the local Lyapunov 
exponents.

What is especially valuable is that the stability analysis allows us to 
decide when we can trust to the results of calculations, even if we do 
not know exact answers. This also permits us do choose between several 
ways of calculations, differring by initial approximations. The general 
guide is always the same: We must select the most stable procedure.

In the present paper, we have considered a variety of examples for which 
accurate numerical data are available. Then, directly comparing the 
latter with our results, we could explicitly demonstrate that the method 
does work. We think that such an explicit demonstration is a necessary 
step before applying the method to more complicated problems for which no 
numerical data are known.

\vskip 1cm

{\Large{\bf Appendix A}}

\vskip 5mm

The integral (77) has the following properties that we have used in the 
process of calculations:
$$
I_{ns}^l(\nu) = I_{sn}^l(\nu) \; , \qquad
I_{ns}^l(0) = \frac{\Gamma(n+l+3/2)}{n!}\;\delta_{ns} \; ,
$$
$$
I_{00}^l(\nu) =\Gamma\left ( l + \frac{3}{2} +\nu\right ) \; , \qquad
I_{0s}^l(\nu) = \frac{\Gamma(l+3/2+\nu)}{s!}\;\varphi_s(-\nu)\; ,
$$
where
$$
\varphi_n(x) \equiv \frac{\Gamma(n+x)}{\Gamma(x)} = x (1+x)(2+x)\ldots
(n-1+x)\; ,
$$
$$
\varphi_0(x) = 1 \; , \qquad \varphi_1(x) = x \; , \qquad 
x\in(-\infty,+\infty) \; ,
$$
$$
\Gamma(-x)\Gamma(x) = -\frac{\pi}{x\sin(\pi x)}\; .
$$
Also,
$$
I_{ns}^l(1) = \left ( 2n + l +\frac{3}{2}\right ) 
\frac{\Gamma(n+l+3/2)}{n!}\; \delta_{sn} - \frac{\Gamma(n+l+3/2)}{(n-1)!}\;
\delta_{s,n-1} - \frac{\Gamma(n+l+5/2)}{n!}\;\delta_{s,n+1} \; .
$$
In general, integral (77) can be presented as the sum
$$
I_{ns}^l(\nu) = \frac{\Gamma(n+l+3/2)}{s!} \sum_{p=0}^n \;
\frac{(-1)^p\; \Gamma(p+l+3/2+\nu)}{p!\;(n-p)!\;\Gamma(p+l+3/2)}\;
\varphi_s(-p-\nu)\; .
$$
Employing these properties, the coefficients in the sequence (76), for 
the case $n=0$, can be simplified to
$$
A_{0l}(\nu) = \frac{\Gamma(l+3/2+\nu)}{(l+3/2)\Gamma(l+3/2)}\; , \qquad
A_{0l}(1) = 1 \; ,
$$
$$
B_{0l}(\nu) = \frac{\nu\Gamma(l+3/2+\nu)}{(l+3/2)\Gamma(l+3/2)}\; , \qquad
B_{0l}(1) = 1 \; ,
$$
$$
C_{0l}(\nu) = \frac{\Gamma^2(l+3/2+\nu)}{(l+3/2)\Gamma(l+3/2)}\;
\sum_{p=1}^\infty \;\frac{\varphi_p^2(-\nu)}{p\;p!\;\Gamma(p+l+3/2)}\; .
$$
The value $\Delta(n,l)$ entering Eq. (78), for $n=0$, writes as
$$
\Delta(0,l) =\frac{1}{8} - \frac{\Gamma(l+5/2)}{2\nu^2} 
\sum_{p=1}^\infty \; \frac{\varphi_p^2(-\nu)}{p\; p!\;\Gamma(p+l+3/2)}\; .
$$

\vskip 1cm

{\Large{\bf Appendix B}}

\vskip 5mm

The integral (86), by employing the replica trick
$$
\ln t = \lim_{\nu\rightarrow 0}\; \frac{t^\nu - 1}{\nu} \; ,
$$
can be expressed through the limit
$$
I_{ns}^l = \lim_{\nu\rightarrow 0}\; \frac{I_{ns}^l(\nu) - I_{ns}^l(0)}{\nu}
$$
involving the integral (77). Taking this limit, we meet the expansion
$$
\frac{\Gamma(x+\nu)}{\Gamma(x)} \simeq 1 + \psi(x)\nu \qquad
(\nu\rightarrow 0) \; ,
$$
in which
$$
\psi(x) \equiv \frac{d}{dx} \ln\Gamma(x) = \sum_{p=0}^\infty\left (
\frac{1}{p+1} - \frac{1}{p+x}\right ) - C = \sum_{p=1}^\infty \;
\frac{x}{p(p+x)} - \frac{1}{x} - C \; ,
$$
$$ 
C\equiv \lim_{n\rightarrow\infty} \left ( \sum_{p=1}^n\; \frac{1}{p} -
\ln n\right ) = 0.577216 \; .
$$
For $n$ integer, one has
$$
\psi(n+1) = \sum_{p=1}^n\; \frac{1}{p} + \psi(1) \; , \qquad 
\psi(1) = -C \; ,
$$
$$
\psi\left ( n +\frac{1}{2}\right ) = \sum_{p=1}^n \frac{2}{2p-1} - 2\ln 2 
- C \; , \qquad \psi\left ( \frac{3}{2}\right ) = 2-\ln 4 - C =0.036490\; .
$$
In the course of taking the limit $\nu\rightarrow 0$, we also need to 
consider the corresponding asymptotic behaviour of the function 
$\varphi_n(-p-\nu)$ defined in Appendix A. This asymptotic behaviour is
\begin{eqnarray}
\varphi_n(-p-\nu) \simeq \left \{ \begin{array}{cc}
(-1)^{p+1}(n-p-1)!\; p!\; \nu & (0\leq p\leq n-1) \\
\nonumber
(-1)^n n!\;\left\{ 1 +\left [\psi(n+1) -\psi(1)\right ]\nu\right\} & (p=n)\; ,
\end{array} \right.
\end{eqnarray}
as $\nu\rightarrow 0$. Using these equalities, for the integral (86), in 
the case of $n=s$, we find
$$
I_{nn}^l =\frac{\Gamma(n+l+3/2)}{n!}\; \psi\left ( n+ l +\frac{3}{2}
\right ) \; .
$$
When $n<s$, then
$$
I_{ns}^l = - \frac{\Gamma(n+l+3/2)}{s!}\; 
\sum_{p=0}^n \; \frac{(s-p-1)!}{(n-p)!} \qquad (n < s ) \; ,
$$
while in the opposite case,
$$
I_{ns}^l = -\frac{\Gamma(n+l+3/2)}{s!} \left [ \;
\sum_{p=0}^{s-1} \; \frac{(s-p-1)!}{(n-p)!} + \sum_{p=0}^{n-s} \;
\frac{(-1)^p}{p!\;(n-p-s)!}\; \psi_{ps}^l\right ] \qquad ( n > s )\; ,
$$
where
$$
\psi_{ps}^l \equiv \psi(p+1) - \psi(p+s+1) - \psi\left ( p+s+l+
\frac{3}{2}\right ) \; .
$$
We also have
$$
I_{np}^l = \lim_{\nu\rightarrow 0}\; \frac{1}{\nu}\; I_{np}^l(\nu) \qquad
(n\neq p) \; ,
$$
$$
I_{n,n-1}^l = - \frac{\Gamma(n+l+1/2)}{(n-1)!} \; , \qquad
I_{n,n+1}^l = - \frac{\Gamma(n+l+3/2)}{n!} \; , \qquad
I_{0s}^l = - \frac{\Gamma(l+3/2)}{s!} \; ,
$$
In this way, we can calculate any coefficient
$$
C_{nl} = \lim_{\nu\rightarrow 0} \; \frac{1}{\nu^2}\; C_{nl}(\nu) \; ,
$$ 
in the sequence (85). In particular, for $n=0$, we get
$$
C_{0l} = \frac{\Gamma(l+3/2)}{l+3/2}\sum_{p=1}^\infty\;
\frac{(p-1)!}{p^2\;\Gamma(p+l+3/2)} \; .
$$

\vskip 1cm

{\Large{\bf Appendix C}}

\vskip 5mm

The integral (102) can be presented as the limit
$$
J_{ns}^l = \lim_{\nu\rightarrow 0}\; \frac{J_{ns}^l(\nu)-J_{ns}^l(0)}{\nu}
$$
involving the integral
$$
J_{ns}^l(\nu) = \int_0^\infty\; t^{2l+2+\nu}\; e^{-t}\; L_n^{2l+1}(t)\;
L_s^{2l+1}(t)\; dt \; .
$$
Substituting in the latter the Laguerre polynomials, we come to the 
expressions similar to those in Appendix A. In this way,
$$
J_{nn}^l(\nu) = \frac{\Gamma(n+2l+2)}{n!}\;\sum_{p=0}^n\;
\frac{(-1)^p\;\Gamma(p+2l+3+\nu)}{p!\; (n-p)!\; \Gamma(p+2l+2)}\;
\varphi_n(-p-1-\nu) \; ,
$$ 
where $\varphi_n(x)$ is the same function as in Appendix A. Also we find
$$
J_{nn}^l(0) = 2(n+l+1)\frac{(n+2l+1)!}{n!} \; , \qquad
J_{nn}^l(-1) = \frac{(n+2l+1)!}{n!} \; .
$$
Taking the limit $\nu\rightarrow 0$, we use the asymptotic equalities
$$
\varphi_n(-p-1-\nu) \simeq (-1)^p\; (n-p-2)!\; (p+1)!\; \nu
$$ for $0\leq p\leq n-2$, and
$$
\varphi_n(-p-1-\nu) \simeq (-1)^n\; (p+1)!\; \left\{ 1 + \left [ 
\psi(p+2) - \psi(p-n+2)\right ]\; \nu \right \} 
$$
for the case of $p=n-1,n$. Here the function $\psi(x)$ is defined in 
Appendix B. With these properties we obtain the expression
$$
J_{nn}^l = \frac{(n+2l+1)!}{n!}\; \left [ 2n+ 1 + 2 (n+l+1)\psi(n+2l+2)
\right ]
$$ defining the coefficients $D_{nl}$ in Eq. (103),
$$
D_{nl} = \frac{n!\; J_{nn}^l}{2(n+l+1)(n+2l+1)!} \; .
$$

\newpage

\begin{center}
{\bf Table captions}
\end{center}

{\bf Table I.} The accuracy of different approximations for the 
ground--state energy $e(0,0)$, for various povers $\nu$ of the potential: 
$\varepsilon_1$ and $\varepsilon_2$ are the percentage errors of the 
renormalized expressions in Eq. (79); $\varepsilon_2^*$ is the error of 
the self--similar approximant (80); and $\varepsilon_{WKB}$ is the error 
of the quasiclassical approximation (73).

\vskip 5mm

{\bf Table II.} The accuracy of approximations for several energy levels 
$e(0,l)$ of the logarithmic potential: $\varepsilon_1$ and $\varepsilon_2$ 
are the percentage errors of the renormalized expressions (89); 
$\varepsilon_2^*$ is the error of the self--similar approximant (90); 
and $\varepsilon_{WKB}$ is the error of the quasiclassical approximation
(83).

\newpage

\begin{center}

{\bf Table I}

\vskip 3mm

\begin{tabular}{|c|c|c|c|c|c|} \hline
$\nu$ & $e(0,0)$ &$\varepsilon_1(\% )$&$\varepsilon_2(\% )$&
$\varepsilon_2^*(\% )$ & $\varepsilon_{WKB}(\% )$ \\ \hline
0.15 & 1.2653 & 0.26 & 0.05 & --0.49 & --0.73 \\
0.50 & 1.5961 & 0.44 & 0.07 & --0.02 & --1.2 \\
0.75 & 1.7450 & 0.39 & 0.06 &   0.08 & --1.0 \\
 1.5 & 2.0121 & 0.08 & 0.01 &   0.04 & --0.26 \\
  2  & 2.1213 & 0    &   0  &    0   &  0   \\
  3  & 2.2765 & 0.33 &--0.05&   0.17 & --0.27 \\
  4  & 2.3936 &  1.3 &--0.43&   0.63 & --1.3 \\
  5  & 2.4924 &  2.6 &--1.6 &    1.2 & --2.6 \\
  6  & 2.5797 &  4.3 &--4.0 &    1.6 & --4.1 \\
  8  & 2.7315 &  8.5 & --16 &    1.0 & --7.1 \\
 10  & 2.8616 &  13  & --50 &  --4.3 & --9.9 \\ \hline
\end{tabular}

\vskip 1cm

{\bf Table II}

\vskip 3mm

\begin{tabular}{|c|c|c|c|c|c|} \hline
$l$ & $e(0,l)$ &$\varepsilon_1(\% )$&$\varepsilon_2(\% )$&
$\varepsilon_2^*(\% )$ & $\varepsilon_{WKB}(\% )$ \\ \hline
0    & 0.52215 & 2.2  & 0.48  &   0.37 & --6.4 \\
1    & 0.82150 & 0.81 & 0.10  & --0.36 & --9.4 \\
2    & 1.0075  & 0.47 & 0.05  & --0.38 & --9.4 \\
3    & 1.1430  & 0.31 & 0.01  & --0.36 & --9.2 \\
4    & 1.2495  & 0.22 &--0.002& --0.33 & --8.9 \\ \hline
\end{tabular}

\end{center}

\end{document}